# Coordinate Space Modification of Fock's Theory. Harmonic Tensors in Quantum Coulomb Problem

Sergei P. Efimov [*]

Bauman Moscow State Technical University,
2-nd Baumanskaya, 5, 105005, Moscow, Russian Federation

We consider Fock's fundamental theory of the hydrogen atom in momentum space which allows a realization of the previously predicted rotation group of a three-dimensional (3D) sphere in four-dimensional (4D) space. We then modify Fock's theory and abandon the momentum space description. To transform and simplify the theory, we use invariant tensor methods of electrostatics in 3D and 4D spaces. We find a coordinate 4D space where the Schrödinger equation becomes the 4D Laplace equation. The transition from harmonic 4D polynomials to original 3D physical space is algebraic and involves derivatives with respect to a coordinate that is interpreted as time. We obtain a differential equation for eigenfunctions in the momentum space and find its solutions. A concise calculation of the quadratic Stark effect is given. The Schwinger resolvent is derived by the method of harmonic polynomials. Vector ladder operators are also considered.

*Keywords:*
Fock's theory, quantum Coulomb problem, harmonic polynomials, ,
the Fourier transform, ladder operators

PACS numbers: 02.30.Em, 03.65.Db, 03.65.Ge(all)

**CONTENTS**

---

[*] *E-mai address:* <u>serg.efimo2012@yandex.ru</u>

# I. INTRODUCTION

## A. *Essence of the problem*

The quantum Coulomb problem, which allows calculating the spectrum of a system of two opposite charges, is still fundamental in quantum theory [1-4]. The names of the founders of twentieth century physics are associated with it: N. Bohr, A. Sommerfeld, V. Pauly, E. Schrödinger and V. Fock. The introduction in the theory of atomic spectra begins with it, and it has been thoroughly studied using methods of the theory of special functions. Due to its simplicity and underlying symmetry- the group SO(4) of rotations in four dimensional space – it is an extremely useful and fine tool of theoretical physics for constructing various concepts [5-7].

 Despite the apparent exhaustive treatment of the quantum Coulomb problem, there are still some questions that have not been fully clarified. In particular, the complexity of calculating the quadratic Stark effect is difficult to understand, even though perturbation used is unmatched in its simplicity in physics [1, 2]. Fock's result is also surprising: why is the SO(4) symmetry realized in the momentum space wrapped into three –dimensional (3D) sphere, with an excursion to the 4D space ?

 Let us recall the background proceeding Fock's accomplishment. Two classical vector integrals, the angular momentum and the Laplace-Runge-Lenz vector, in quantum mechanics correspond to vector operators that commute with the energy operator, i.e., with the Hamiltonian. An analysis of their commutators[1] carried out in [8] shows that they generate a Lie algebra (a linear space with a commutation operation) coinciding with a Lie algebra of operators of small (infinitesimal) rotations in 4D space [1,3].

 For physicists, the correspondence means that some transformation of variables and operators maps the original quantum Coulomb problem into the problem of free motion of a particle over a 3D sphere embedded in a 4D space. The energy operator is then invariant under rotations of the 3D sphere . This is reminiscent of the remarkable effect of Lewis Carroll's soaring grin of the Cheshire cat.

 Fock' approach struck contemporaries [3, 9-11]. The starting point in his theory is integral Schrödinger's equation (**SE**) in momentum space. The space can be considered as 3D plane in a 4D space. Fock then wraps 3D plane into a 3D sphere using stereographic projection, known since antiquity [12] as convenient transformation of a globe into a flat map[13]. (Fock's globe is three-dimensional as is the map.) At the same time, additionally, Fock surmises the factor for psi functions (for solutions) such that original integral equation turns into an equation for spherical functions on 3D sphere (to be distinguished from functions on two-dimensional sphere).

---

[1] The Laplace-Runge-Lenz operator is first multiplied by *n* [3].

This equation rarely used in physics but well-known in the theory of special functions, is invariant under rotations in 4D space. Fock doesn't explain the physical meaning of transformation he found [9-11]. As a result, the fundamental questions remain: why is the SO(4) symmetry realized in wrapped momentum space rather than in the coordinate space, and how the electron "learned about stereographic projection?"

In this paper, Fock's approach is modified by the inverse 4D Fourier transform which is applicable to eigenfunctions extended harmonically to the 4D momentum space[2]. This is transition to a new coordinate space, because Fock transforms **SE** into 4D momentum space. In the 4D coordinate space, the modified **SE** is the 4D Laplace equation, and the eigenfunctions are harmonic 4D tensors, i.e., homogeneous polynomials. Eigenfunctions are harmonic tensors now, i.e., uniform polynomials. Their projections (i.e., contractions with numerical tensors) are solid spherical 4D functions.

The final transition to solutions of the original **SE** in Fock's theory turns out to be simple algebra with the help of differentiation. The SO(4) symmetry is thus realized in a coordinate space that, in a certain sense, is 'closer here' than the space found by Fock. Although the structure of the functions is preserved, the inverse transformation is remarkably simple, which is of importance for theoretical concepts.

In this paper, spherical functions are replaced by polynomials that have been well known in electrostatics, since the time of Maxwell, and are associated with multipole moments [14-24]. In particular, instead of solid functions $r^l Y_{l,m}(\theta,\varphi)$ tensor polynomials $x_i$ and $(3x_i x_k - r^2 \delta_{ik})$ generating the dipole and quadrupole moments are used for l= 1, 2, and octupole polynomial is used for l=3, and so on. Spherical functions are convenient in scattering problems. Polynomials are preferable in calculation with differential operators and are therefore introduced here from the very beginning. Spherical coordinates $\theta,\varphi$ and $\theta,\varphi,\phi$ are not involved.

In physics, dipole and quadrupole moments typically appear because the fundamental concepts of physics are associated with them. But the use of invariant polynomial tensors in Cartesian coordinates, as shown in a number of recent studies, is preferable and simplifies the fundamental scheme of calculations [25-31]. In Section **3**, the rules for using harmonic symmetric tensors are demonstrated that directly follow from their properties. This rules are naturally reflected in the theory of special functions, but are not always obvious, even though the group properties are general [32]. At any rate, we recall the main property of harmonic tensors: the trace over any pair of indices vanishes [23, 33].

---

[2] Applying the inverse (3D) Fourier transform would mean a return to the original SE in the physical space

We here select those properties of tensors that not only make analytic calculations more compact and reduce 'the number of factorials' but also allow correctly formulating some fundamental questions of the theory. The advantage of using such tensors in the perturbation theory is demonstrated in Sections 4 and 9 in calculating the quadratic Stark effect and in deriving the Schwinger resolvent. In Section 7, we present the derivation of the integral equation obtained by Fock with reference to the theory of special functions, using methods of electrostatics in 3D and 4D spaces. In Section 8, we derive the vector ladder operators in the Coulomb problem and show their relation to inversion with respect to a sphere.

To modify the Fock theory, it is convenient to use harmonic tensors in moving to the 4D coordinate space, because, as shown below, their form remains invariant under rather complex transformations. Of course, the SO(4) symmetry is inherent in invariant 4D tensors. The final algebraic transition to the physical space singles out one coordinate, which can be called complex time. This coordinate is then multiplied by the imaginary unit $i$ (going into real time) and equated to the length of the radius vector $|\mathbf{r}|$, thus being 'absorbed' by the formula. The SO(4) symmetry is thus concealed in the original solution of the SE.

Strictly speaking, the transformation found in this paper is not identical to Fock's theory. There three fundamental variations that simplify the transformations used. The first is the extension from the Fock sphere into 4D space, when harmonic tensors are found. The second is the use of spatial 4D inversion instead of the stereographic projection of the sphere. The third variation amounts to discarding delta functions in the final transformation. The SO(4) symmetry is preserved each time, which leads to the desired result: the SE eigenfunctions are algebraically related to harmonic 4D tensors. We note that the state correspondence found in this paper is generated by the physical symmetry of the problem and is not known in theory of the Laguerre and the Gegenbauer polynomials [32].

**B.** *Harmonic polynomials instead of spherical functions*

The Schrödinger equation (**SE**) for eigenfunctions, using atomic units (unit of energy being $\frac{Z^2 me^4}{\hbar^2}$ and the unit of length being Bohr's radius $a_B = \frac{\hbar^2}{Zme^2}$), has the form :

$$(-\frac{1}{2}\Delta - \frac{1}{r})\Psi_{nl} = -\frac{1}{2n^2}\Psi_{nl} . \qquad (1)$$

In what follows, it is convenient to reduce the orbits of all radii [1], i.e., change the radius vector for each eigenfunction as $\mathbf{r}' = \frac{\mathbf{r}}{n}$. Eq. (1) then takes a deceptively simple form,

$$(-\Delta + 1)\Psi_{nl} = \frac{2n}{r}\Psi_{nl} , \qquad (2)$$

where $r$ is the modulus of the vector **x**. Eigenfunctions in momentum representation then have scaled argument $\mathbf{p}' = n\mathbf{p}$.

In this paper, the angular factor is not taken in the form of spherical functions but as a combination of solid spherical functions, i.e., a harmonic degree-$l$ polynomial $P_l(\mathbf{x})$ homogeneous in coordinates with the properties

$$\Delta P_l(\mathbf{x}) = 0, \quad P_l(c\mathbf{x}) = c^l P_l(\mathbf{x}), \quad (\mathbf{x}\nabla) P_l(\mathbf{x}) = l P_l(\mathbf{x}). \qquad (3)$$

The last equality is Euler's theorem which is valid for all homogeneous functions (not only polynomials but also, for example, the factor 1/r and its powers). The operator that arises here has an interpretation that allows calling it the *angular momentum modulus operator*:

$$\hat{l} = (\mathbf{x}\nabla). \qquad (4)$$

Solutions of Eq. (2) and similar equations can be represented as $P_l(\mathbf{x})F(r)$. When substituted into Eq. (2) and others equations containing the Laplace operator, properties (3) give rise to the equation for the radial function because the factor $P_l(\mathbf{x})$ persists and can be canceled:

$$\Delta[P_l(\mathbf{x})F(r)] = [\Delta P_l(\mathbf{x})]F(r) + 2[\nabla P_l(\mathbf{x})][\nabla F(r)] + P_l(\mathbf{x})\Delta F(r) =$$
$$= (2l)P_l(\mathbf{x})\frac{1}{r}\frac{\partial F(r)}{\partial r} + P_l(\mathbf{x})\frac{1}{r}\frac{\partial^2}{\partial r^2}[rF(r)] \qquad (5)$$

This transformation, after the substitution in Eq. (2), immediately leads to the Gauss function[3]

$$F(\alpha,\beta,r) = 1 + \frac{\alpha}{\beta}\frac{r}{1!} + \frac{\alpha(\alpha+1)}{\beta(\beta+1)}\frac{r^2}{2!} + \frac{\alpha(\alpha+1)(\alpha+2)}{\beta(\beta+1)(\beta+2)}\frac{r^3}{3!}\ldots. \qquad (6)$$

For a discrete spectrum, parameter $\alpha$ is the negative integer, which converts the series into polynomial, and the eigenfunctions (taking linear scaling of the argument $\mathbf{x}$ into account) in the Coulomb problem are considered in the form [1]

$$P_l(\mathbf{x})F(-k, 2l+2, 2r)\exp(-r), \quad n = l+k+1. \qquad (7)$$

The number $k$ is the degree of the second factor in Eq. (7), and hence, the degree of the entire polynomial is n-1. The normalization here and below is unimportant, unless
especially stipulated.

If there is no preferred direction and the spectrum is independent of the quantum number $m$ then the entire set of eigenfunctions with different values of $m$ can be considered simultaneously. For the transformations considered below, it is convenient to use rank-$l$ tensors invariant under the rotation group SO(3) that were developed in electrostatics. In particular, for the quadruple state, the solution

---
[3] The degenerate hyper-geometric function $_1F_1(r)$

has the form

$$(3x_i x_k - r^2 \delta_{ik})F(-k,6,2r)\exp(-r), \quad n = 3+k. \tag{8a}$$

For the octupole state, the solution is

$$3(5x_i x_k x_l - x_i \delta_{kl} - x_l \delta_{ik} - x_k \delta_{li})F(-k,8,2r)\exp(-r), \quad n = 4+k. \tag{8b}$$

For the dipole state, the solution is obvious: $x_i F(-k,4,2r)\exp(-r)$. The general form of an invariant harmonic tensor and its properties that are useful in quantum the Coulomb problem, are considered in Section 3.

## II. FOCK'S THEORY

Schrödinger equation (2) (with $\hbar = 1$) when moving to the momentum representation

$$\Psi_{nl}(\mathbf{x}) = \frac{1}{(2\pi)^3}\int a_{nl}(\mathbf{p})e^{i(\mathbf{p}\mathbf{x})}d^3\mathbf{p}, \tag{9}$$

contains a convolution with respect to momenta. Because potential 1/r pass in $4\pi/|\mathbf{p}^2|$, the **SE** in the momentum space is nonlocal:

$$(|\mathbf{p}^2|+1)a_{nl}(\mathbf{p}) - \frac{2n}{2\pi^2}\int\frac{a_{nl}(\mathbf{p}')d^3\mathbf{p}'}{|\mathbf{p}-\mathbf{p}'|^2} = 0. \tag{10}$$

First step of the theory is to multiply the function $a_{nl}(\mathbf{p})$ (done without an explanation) by $(1+\mathbf{p}^2)^2$. The second step is to wrap the 3D plane into 3D sphere (with four coordinates $(\boldsymbol{\xi}, \xi_0)$; see Fig.1).[4]

Fig. 1 shows that tangent of the slope of the projecting (red) straight line is

$$\tan\varphi = \frac{1}{|\mathbf{p}|}.$$

---

[4] In [3], the sign of $\xi_0$ is reversed

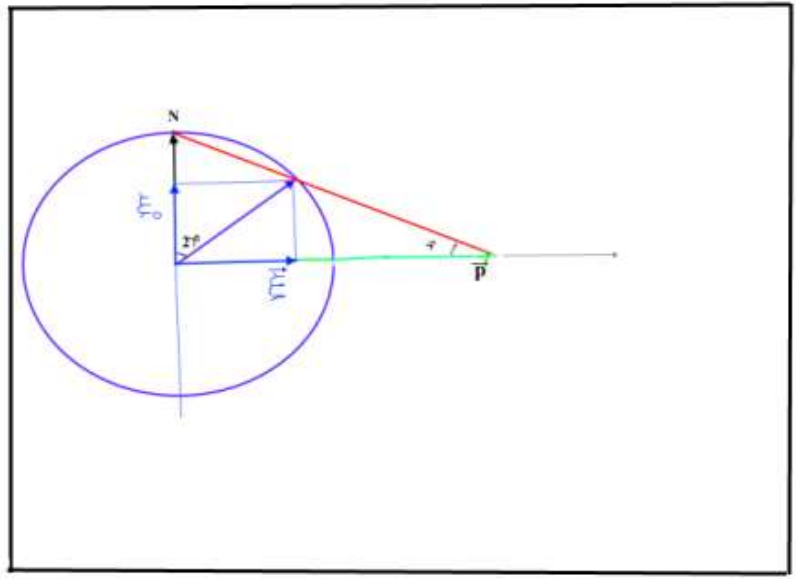

**FIG.** 1. Stereographic projection of a 3D **p-** plane
onto a unit radius 3D sphere

Hence, it follows that

$$|\boldsymbol{\xi}| = \sin 2\varphi = \frac{2|\mathbf{p}|}{(1+\mathbf{p}^2)}, \quad \boldsymbol{\xi} = \frac{2\mathbf{p}}{(1+\mathbf{p}^2)},$$
$$\xi_0 = \cos 2\varphi = \frac{(\mathbf{p}^2-1)}{(\mathbf{p}^2+1)}, \quad \boldsymbol{\xi}^2 + \xi_0^{\,2} = 1. \tag{11}$$

The stereographic projection doubles the tilt angle $\varphi$, and this is essential effect it produces. The flat drawing correctly reflects the 4D transformation.

In the new variables, with Fock's factor taking into account, the eigenfunction becomes

$$b_{nl}(\boldsymbol{\xi}, \xi_0) = (\mathbf{p}^2+1)^2 a_{nl}(\mathbf{p}). \tag{12}$$

It is essential as well that the projection to be given by a conformal transformation. The angles between intersecting curves are preserved. The metric on the sphere in the momentum-space coordinates (of the **p-**plane) is expressed as

$$\frac{4}{(\mathbf{p}^2+1)^2}(d\mathbf{p})^2. \tag{13}$$

Hence, the contraction coefficient for elements of the **p**-space is $(1+\mathbf{p}^2)/2$.
The volume element in formula (10) is expressed in terms of the 3D surface element:

$$d^3\mathbf{p} = \frac{1}{8}(1+\mathbf{p}^2)^3 dS_3. \tag{14}$$

The kernel of integral can be (very fortunately but not obviously) transformed as

$$\frac{1}{|\mathbf{p}-\mathbf{p'}|^2} = \frac{2}{(\mathbf{p}^2+1)} \frac{1}{[(\boldsymbol{\xi}-\boldsymbol{\xi'})^2+(\xi_0-\xi'_0)^2]} \frac{2}{(\mathbf{p'}^2+1)}, \qquad (15)$$

that doesn't follow from the conformal property. In particular, equality (15) is greatly simplified for two opposite points of the sphere: $|\mathbf{p}|=|\mathbf{p'}|=1$, $\xi_0=\xi_0'=0$, $\mathbf{p}=-\mathbf{p'}$, $\boldsymbol{\xi}=-\boldsymbol{\xi'}$.

Now, substituting Eqs. (12), (14) and (15) into integral Eq. (10), we obtain

$$b_{nl}(\boldsymbol{\xi},\xi_0) - \frac{n}{2\pi^2}\int \frac{b_{nl}(\boldsymbol{\xi'},\xi_0')dS_{\xi'}}{[(\boldsymbol{\xi}-\boldsymbol{\xi'})^2+(\xi_0-\xi_0')^2]} = 0, \qquad (16)$$

where, as can be seen from the figure, the surface element on the unit 4D sphere with the volume is, $2\pi^2$

$$dS_{\xi'} = \frac{d\xi_1'd\xi_2'd\xi_3'}{\xi'_0} = \frac{dV_{\xi'}}{\xi'_0}. \qquad (17)$$

(Integration over flat space $\xi_0' = 0$ is convenient for calculations.)

Next, Fock refers to the theory of spherical functions in 4D space. However, any spherical function and a sum of such functions with fixed index (n-1) that corresponds to the value of $n$ in the original SE, can be substituted into Eqn (16),[5] and hence the equation does not fix quantum numbers $l$ and $m$. The property of being conformal is important here. A rotation on the sphere corresponds to a rotation through the same angle in the momentum and physical spaces, and therefore a function with the factor $Y_{lm}(\theta,\varphi)$ goes into eigenfunction with the same factor. If the factor is an invariant tensor, the same tensor is contained in the original function.

Eq. (16), as we show below, naturally arises in 3D and 4D electrostatics. This analogy is discussed in Section 7.

### III. HARMONIC and MULTIPOLE TENSORS

#### A. *Multipole tensors in 3D electrostatics*

We recall that multipole potentials arise when the potential of point charge is expanded in powers of coordinates $x_{0i}$ of radius vector $\mathbf{r}_0$ ('Maxwell poles') [17, 18]. For the $1/|\mathbf{r}-\mathbf{r}_0|$ potential, we have

$$\frac{1}{|\mathbf{r}-\mathbf{r}_0|} = \sum_l (-1)^l \frac{(\mathbf{r}_0\nabla)^l}{l!}\frac{1}{r} = \sum_l \frac{x_{0i}...x_{0k}}{l!r^{2l+1}}\mathbf{M}_{i...k}^{(l)}(\mathbf{r}) = \sum_l \frac{(\mathbf{r}_0^{\otimes l},\mathbf{M}_{[i]}^{(l)})}{l!r^{2l+1}}, \qquad (18)$$

---
[5] In [3], the value of n is given

where following notations are used. For the $l$th tensor power of the radius-vector, we wright

$$x_{0i}...x_{0k} = \mathbf{r}_0^{\otimes l},$$

and for a symmetric harmonic tensor of rank $l$,

$$\mathbf{M}_{i...k}^{(l)}(\mathbf{r}) = \mathbf{M}_{[\mathbf{i}]}^{(l)}(\mathbf{r}). \tag{19}$$

Tensor from Eq. (19) is homogeneous harmonic polynomial (with properties (3)). Contraction over any two indices (when the scalar product of two gradients become the Laplace operator) is zero. If tensor of Eq. (19) divided by factor $r^{2l+1}$, then *multipole* harmonic tensor arises,

$$\frac{\mathbf{M}_{i...k}^{(l)}(\mathbf{r})}{r^{2l+1}} = \frac{\mathbf{M}_{[\mathbf{i}]}^{(l)}(\mathbf{r})}{r^{2l+1}}, \tag{20}$$

which is also a homogeneous harmonic function with negative homogeneity degree $-(l+1)$.

The theory typically involves the dipole, quadrupole, and octupole tensors entering Eq. (8). The higher-rank ones are used less frequently due to their presumed complexity. However, if we recall the obvious property of contraction

$$(2l-1)!!(\mathbf{r}_0^{\otimes l}, \mathbf{M}_{[\mathbf{i}]}^{(l)}(\mathbf{r})) = \left(\mathbf{M}_{[\mathbf{i}]}^{(l)}(\mathbf{r}_0), \mathbf{M}_{[\mathbf{i}]}^{(l)}(\mathbf{r})\right), \tag{21}$$

then, in calculating the potential amplitudes, we do not have to use spherical functions of tensors of Eq. (19): it suffice to calculate simple power-law moments. This rule is so important in calculations with tensors that it can be formulated as a theorem of electrostatics.

**B.** *Theorem on power-law equivalent moments in electrostatics*

Let $\rho(\mathbf{x})$ be a distribution of charge. When calculating a multipole potential, power-law moments can be used instead of spherical functions (or instead of harmonic tensors of Eq. (19)):

$$\int \frac{\rho(\mathbf{r}_0)}{|\mathbf{r}-\mathbf{r}_0|}dV_{\mathbf{r}_0} = \sum_l \int \rho(\mathbf{r}_0)\mathbf{M}_{[\mathbf{i}]}^{(l)}(r_0)dV_{\mathbf{r}_0} \frac{\mathbf{M}_{[\mathbf{i}]}^{(l)}}{(2l-1)!!l!r^{2l+1}} = \sum_l \int \rho(\mathbf{r}_0)\mathbf{r}_0^{\otimes l}dV_{\mathbf{r}_0} \frac{\mathbf{M}_{[\mathbf{i}]}^{(l)}}{l!r^{2l+1}}. \tag{22}$$

*Example (1):*
For octupole moment, instead of integral in the third term in the first sum in (22)

$$\int \rho(\mathbf{r})3(5x_i x_k x_l - x_i\delta_{kl} - x_k\delta_{li} - x_l\delta_{ki})dV_{\mathbf{r}}$$

we can take 'short' integral in the second sum,

$$15\int \rho(\mathbf{r}_0)x_{0i}x_{0k}x_{0l}dV_{\mathbf{r}_0}$$

and substitute it into the second sum in Eq. (22). what gives the last item. The moments are different, but the potentials are the same.

## C. *Formula for a harmonic tensor*

Formula for the tensor was considered in [26, 27] using a ladder operator. Here, it is derived using the Laplace operator. The first term in the formula, as is easy to see from (18), is equal to

$$\mathbf{M}_{[\mathbf{i}]}^{(l)}(\mathbf{r}) = (2l-1)!! \, x_{i_1} \ldots x_{i_l} + \ldots = (2l-1)!! \, \mathbf{r}^{\otimes l} + \ldots \quad (23)$$

The remaining terms are obtained by repeatedly applying the Laplace operator and multiplying by an even power of the modulus $|\mathbf{r}|^{2k}$. The coefficients are easy to determine by substituting Eq. (23) in the Laplace equation (See Appendix A). As a result, we have

$$\mathbf{M}_{[\mathbf{i}]}^{(l)}(\mathbf{r}) = (2l-1)!! \, \mathbf{r}^{\otimes l} - \frac{(2l-3)!!}{1!2^1} r^2 \Delta \mathbf{r}^{\otimes l} + \frac{(2l-5)!!}{2!2^2} r^4 \Delta^2 \mathbf{r}^{\otimes l} - \frac{(2l-7)!!}{3!2^3} r^6 \Delta^3 \mathbf{r}^{\otimes l} + \ldots \quad (24)$$

This form is useful for applying the differential operators of quantum mechanics and electrostatics to it. The differentiating in (24) generates products of the Kronecker symbols.

*Example (2):*

$$\Delta x_i x_k = 2\delta_{ik}, \quad \Delta x_i x_k x_l = 2(\delta_{ik} x_l + \delta_{kl} x_i + \delta_{li} x_k),$$
$$\Delta\Delta x_i x_k x_l x_m = 8(\delta_{ik}\delta_{lm} + \delta_{il}\delta_{km} + \delta_{im}\delta_{kl}). \quad (25)$$

The last quality can be verified using the contraction with $i = k$. It is convenient to write the differentiation formula in terms of the symmetrization operation. A symbol for it was proposed in [27], with the sum taken over all independent permutations of indices.[6]

$$\frac{\Delta^k \mathbf{r}^{\otimes l}}{k! 2^k} = \ll \delta_{[..]}^{\otimes k} \mathbf{r}^{\otimes (l-2k)} \gg . \quad (26)$$

As a result, we obtain the formul
$$\mathbf{M}_{[\mathbf{i}]}^{(l)}(\mathbf{r}) =$$
$$= (2l-1)!! \, \mathbf{r}^{\otimes l} - (2l-3)!! \, r^2 \ll \delta_{[..]}^{\otimes 1} \mathbf{r}^{\otimes (l-2)} \gg + (2l-5)!! \, r^4 \ll \delta_{[..]}^{\otimes 2} \mathbf{r}^{\otimes (l-4)} \gg \ldots \quad (27)$$

where symbol $\delta_{[..]}^{\otimes k}$ is used for a tensor power of the Kronecker symbol and the conventional symbol [..] is used for the two subscripts that changed under symmetrization.

We follow [26] in finding the relation between the tensor and solid spherical functions. We use two unit vectors: vector $\mathbf{n}_z$ directed along z-axis and complex vector $(\mathbf{n}_x \pm \mathbf{n}_y) = \mathbf{n}_\pm$. Contraction with their powers gives required relation

$$(\mathbf{M}^{(l)}_{[\mathbf{i}]}, \mathbf{n}^{\otimes(l-m)} \mathbf{n}_\pm^{\otimes m}) = (l-m)!(x\pm iy)^m r^{(l-m)} \frac{d^m}{dt^m} P_l(t)\bigg|_{t=z/r},$$

where $P_l(t)$ is a Legendre polynomial.

---
[6] The factor k! lost in [26].

### D. *Harmonic 4-D tensors*

The potential of a point charge[7] in 4D space is equal to $1/r^2$. From the expansion of the point-charge potential of form (18), we obtain the multipole 4-D potential

$$\frac{\mathcal{M}^{(n)}_{i...k}(\mathbf{r})}{r^{2n+2}} = (-1)^n \nabla_i ... \nabla_k \frac{1}{r^2}. \tag{28}$$

The harmonic tensor in the nominator has a structure similar to Eq. (27). Its contraction with respect to any two indices must vanish. The dipole and quadruple 4-D tensors, as follows from Eq. (28), are expressed as

$$\mathcal{M}^{(1)}_i(\mathbf{r}) = 2x_i, \quad \mathcal{M}^{(2)}_{ik}(\mathbf{r}) = 2(4x_i x_k - \delta_{ik}). \tag{29}$$

The leading term of the expansion, as can be seen from (28), is equal to

$$\mathcal{M}^{(n)}_{[\mathbf{i}]}(\mathbf{r}) = (2n)!! x_{i_1}...x_{i_n} + ... = (2n)!! \mathbf{r}^{\otimes n} + ... . \tag{30}$$

Using the method described in Section 3.3, we obtain two representations (because differentiation rule of Eq. (26) does not change):

$$\mathcal{M}^{(n)}_{[\mathbf{i}]}(\mathbf{r}) = (2n)!! \mathbf{r}^{\otimes l} - \frac{(2n-2)!!}{1!2^1} r^2 \Delta \mathbf{r}^{\otimes n} + \frac{(2n-4)!!}{2!2^2} r^4 \Delta^2 \mathbf{r}^{\otimes n} + ... . \tag{31}$$

$$\mathcal{M}^{(n)}_{[\mathbf{i}]}(\mathbf{r}) = (2n)!! \mathbf{r}^{\otimes n} - (2n-2)!! r^2 \ll \delta^{\otimes 1}_{[..]} \mathbf{r}^{\otimes(n-2)} \gg + (2n-4)!! r^4 \ll \delta^{\otimes 2}_{[..]} \mathbf{r}^{\otimes(n-4)} \gg ... . \tag{32}$$

We recall the doubled factorial

$$(2n)!! = 2^n n!, \tag{33}$$

which is essential in Eq. (32).

Four-dimensional tensors are structurally simpler than 3D tensors. But the SE requires that eigenfunction be multiplied by a 3D tensor. This forces us to use not the 4D tensor itself but its special projection, which loses 4D invariance as 4D indices are replaced with 3D indices.

---

[7] We define unit of charge in terms of the point potential

## E. *Decomcoordinate of polynomials in terms of harmonic functions*

In perturbation theory, it is necessary to expand the source in terms of spherical functions. If the source is a polynomial, for example, when calculating the Stark effect, then the integrals are standard, but cumbersome. When calculating with the help of invariant tensors, the expansion coefficients are simplified, and there is then no need to integrals. It suffices, as we show here, to calculate contractions that lower the rank of the tensors under consideration.

Instead of integrals, we use the operation of calculating the trace $\hat{Tr}$ of a tensor over two indices. The following rank reduction formula is useful.[8]

$$\hat{Tr} \ll \delta_{ik} M^{(l)}_{ik[\mathbf{m}]} \gg = (2l+3) M^{(l-2)}_{[\mathbf{m}]} . \tag{34}$$

If the brackets contain several Kronecker symbols, the following relation formula holds:

$$\hat{Tr} \ll \delta_{i_1 p_1} ... \delta_{i_k p_k} M^{(l)}_{i_1 p_1 [\mathbf{m}]} \gg = (2l+2k+1) \ll \delta_{i_2 p_2} ... \delta_{i_k p_k} M^{(l-2)}_{[\mathbf{m}]} \gg . \tag{35}$$

Calculating the trace reduces the number of Kronecker symbols by one, and the rank of the harmonic tensor on the right-hand side of Eq. (35) decreases by two. Repeating the calculation of the trace $k$ times eliminates the Kronecker symbols:

$$\hat{Tr}_1 ... \hat{Tr}_k \ll \delta_{i_1 p_1} ... \delta_{i_k p_k} M^{(l)}_{[\mathbf{m}]} \gg = (2l+2k+1)!! M^{(l-2k)}_{[\mathbf{m}]} \tag{36}$$

Applying these rules allows decomposing the tensor $x_i...x_p$ with respect to the harmonic ones.

In the perturbation theory, even the third approximation often considered good. We present the decomposition of the tensor power up to the rank $l=6$:

- $l=2$    $3 x_i x_k = M^{(2)}_{ik} + r^2 \delta_{ik}$
- $l=3$    $5!! x_i x_k x_m = \mathrm{M}^{(3)}_{ikm} + 3r^2 \ll \delta_{ik} x_m \gg$
- $l=4$    $7!! x_i x_k x_l x_m = \mathrm{M}^{(4)}_{iklm} + 5 r^2 \mathrm{M}^{(2)}_{ik} \delta_{lm} \gg + 7 r^4 \ll \delta_{ik} \delta_{lm} \gg$
- $l=5$    $9!! x_i x_k x_l x_m x_n = \mathrm{M}^{(5)}_{iklmn} + 7 r^2 \ll \mathrm{M}^{(3)}_{ikl} \delta_{mn} \gg + 3 \times 9 r^4 \ll \delta_{ik} \delta_{lm} x_n \gg$

---

[8] The symmetrization operation is not associative. Here, it is applied to tensor as a whole. Formula (27) can be applied to the left-hand side of (34) only after symmetrization.
In particular $\ll \delta_{[..]} \ll \delta_{[..]} x_m \gg \gg = 2 \ll \delta_{[..]} \delta_{[..]} x_m \gg$

- $l=6$

$$11!! x_i x_k x_l x_m x_n x_p = \mathbf{M}^{(6)}_{iklmnp} + 9r^2 \ll \mathbf{M}^{(4)}_{iklm}\delta_{np} \gg + 5\times 11 r^4 \ll \mathbf{M}^{(2)}_{ik}\delta_{lm}\delta_{np} \gg +$$
$$+ 9\times 11 r^6 \ll \delta_{ik}\delta_{lm}\delta_{np} \gg \tag{37}$$

To derive formulas, it is useful to calculate the trace. The formula for $l=6$ then implies the formula for $l=4$. The trace is calculated using rule of Eq. (35). For even values of $l$, the last term in Eq. (37) has the form

$$\frac{(2l-1)!!}{(l+1)!!} . \tag{38}$$

When averaging over the directions of the vector $x_i$, i.e., when integrating over the unity sphere, it is convenient to use a simple relation for even values of $l$ [21],

$$\frac{1}{4\pi}\int_{|\mathbf{x}|=1} x_{i_1}...x_{i_l} dS_{\mathbf{x}} = \frac{1}{(2l+1)!!} \ll \delta_{i_1 i_2}...\delta_{i_{(l/2-1)} i_{l/2}} \gg, \tag{39}$$

which is easy to verify by calculating the trace (see Eq. (36)). In particular, the most commonly used averaging in physics is obviously the one with $l=2$ [1,21]:

$$\frac{1}{4\pi}\int_{|\mathbf{x}|=1} x_i x_k dS_{\mathbf{x}} = \frac{1}{3}\delta_{ik} . \tag{40}$$

Also useful is the frequently occurring contraction over all indices without integration,

$$(M^{(l)}_{[\mathbf{i}]}, M^{(l)}_{[\mathbf{i}]}) = M^{(l)}_{i...k}(\mathbf{x}) M^{(l)}_{i...k}(\mathbf{x}) = \frac{(2l)!}{2^l} r^{2l} ,$$

which arises when normalizing the states.

In *four-dimensional* space, formulas preserve their simplicity. The position of indices in Eq. (31) are unique and can therefore be omitted. Instead of Eq. (35), we then have

$$\hat{T}r \ll \delta^{\otimes k}_{[..]} \mathcal{M}^{(n)}_{[..][\mathbf{m}]} \gg = (2l+2k+2) \ll \delta^{\otimes(k-1)}_{[..]} \mathcal{M}^{(n-2)}_{[\mathbf{m}]} \gg . \tag{41}$$

The decomposition of tensor powers of a vector is also compact in four dimensions:
- n=2   $\quad 4!! x_i x_k = \mathcal{M}^{(2)}_{ik} + 2r^2 \delta_{ik}$
- n=3   $\quad 6!! x_i x_k x_m = \mathcal{M}^{(3)}_{ikm} + 8r^2 \ll \delta_{ik} x_m \gg$
- n=4   $\quad 8!! x_i x_k x_l x_m = \mathcal{M}^{(4)}_{iklm} + 6r^2 \ll \mathcal{M}^{(2)}_{ik}\delta_{lm} \gg + 16r^4 \ll \delta_{ik}\delta_{lm} \gg$

- n=5    $10!! x_i x_k x_l x_m x_n = \mathcal{M}^{(5)}_{iklmn} + 8r^2 \ll \mathcal{M}^{(3)}_{ikl} \delta_{mn} \gg + 10 \times 8 r^4 \ll \delta_{ik} \delta_{lm} x_n \gg$
- n=6

$$12!! x_i x_k x_l x_m x_n x_p = \mathcal{M}^{(6)}_{iklmnp} + 10 r^2 \ll \mathcal{M}^{(4)}_{iklm} \delta_{np} \gg + 12 \times 6 r^4 \ll \mathcal{M}^{(2)}_{ik} \delta_{lm} \delta_{np} \gg +$$
$$+ 24 \times 10 r^6 \ll \delta_{ik} \delta_{lm} \delta_{np} \gg$$

(42a)

When using the tensor notation (with indices suppressed), the last equality becomes

$12!! \mathbf{x}^{\otimes 6} =$

$= \mathcal{M}^{(6)}_{[\mathbf{i}]} + 10 r^2 \ll \mathcal{M}^{(4)}_{[\mathbf{i}]} \delta_{[..]} \gg + 12 \times 6 r^4 \ll \mathcal{M}^{(2)}_{[\mathbf{i}]} \delta^{\otimes 2}_{[..]} \gg + 24 \times 10 r^6 \ll \delta^{\otimes 3}_{[..]} \gg$   (42b)

We note that the derivation of coordinate-wise Eqs. (37) and (41) by integrating spherical functions is rather cumbersome. Decomposition of higher powers is not more difficult using contractions over two indices.

## IV. STARK'S QUADRATIC EFFECT

### A. *Method of the first correction*

To demonstrate the effectiveness of calculations with polynomials ( and with harmonic tensors), we consider quadratic the Stark effect in the case where the linear effect is absent. Appropriate formula was obtained in [34-38] by passing to parabolic coordinates. A modern presentation can be found in [1, 39,40]. The **SE** then separates and two sets of orthogonal functions arise. The perturbation matrix automatically become diagonal which immediately leads to the well known formula for the dipole moment proportional to the difference between parabolic quantum numbers $(n_1 - n_2)$.

If $n_1 = n_2 = m$ and there is no linear effect, then quadratic effect must be calculated. In parabolic coordinates, the calculation of the second approximation is not so obvious and requires some effort [1]. The final formula is as follows. The dipole moment $d_z$ is proportional to the electric field $\mathcal{E}$:

$$d_z = \frac{n^4}{8}(17 n^2 - 9 m^2 + 19)\mathcal{E},  \qquad (43)$$

where $n$ and $m$ are the quantum numbers of the hydrogen atom. For the ground state with $n = 1$ and $m = 0$, the electric polarizability is $\frac{9}{2} a_B^3$, which is a fundamental law of nature.

Here, we consider the calculation of the Stark quadratic effect using a technique known in physics, when the symmetry properties (for example, the unitarity of scattering matrix) allow obtaining the next approximation algebraically in terms of preceding ones [25]. The calculation of matrix elements is not required.

When expanded in powers of $\mathcal{E}$, the original SE for the first correction to the state $\Psi_1$ has the form

$$(\frac{1}{2}\Delta + E_0 + \frac{1}{r})\Psi_1 = \mathcal{E} z \Psi_0 , \tag{44}$$

where the electron charge is negative. We are interested in the second approximation to energy $E_2$ in the case where the first correction vanishes:

$$E_1 = (\Psi_0, z E \Psi_0) = 0 . \tag{45}$$

Then, the second correction to the energy is calculated linearly in terms of the first correction:

$$E_2 = \frac{1}{2}(\Psi_1, \mathcal{E} z \Psi_0) .$$

If the function $\Psi_0$ is not normalized, the right-hand side must be divided by the norm squared:

$$E_2 = \frac{1}{2}(\Psi_1, \mathcal{E} z \Psi_0) \frac{1}{(\Psi_0, \Psi_0)} . \tag{46}$$

We note that Eq. (44) gives a non-unique solution for function $\Psi_1$. We can always add $\Psi_0$ to it, because the left-hand side of Eq. (44) vanishes for $\Psi_0$. But this does not change the quadratic correction of Eq. (46), because the linear correction is equal to zero. Thus, the problem reduces to calculating the first correction $\Psi_1$. In this case the polynomial decomposition reduces the number of computations. As an illustration, we consider two problems that are simple in their original formulation.

**B. *Ground state n=1***

For $\Psi_0 = \exp(-r)$, Eq.(44) takes the form

$$(\frac{1}{2}\Delta - \frac{1}{2} + \frac{1}{r})\Psi_1 = \mathcal{E} z e^{-r} . \tag{47}$$

The solution in the form of a polynomial multiplied by an exponential, is easy to guess:

$$\Psi_1 = -(1 + \frac{r}{2})\mathcal{E} z e^{-r} .$$

When substituting this into Eq. (47), it is helpful to use Euler's theorem. Simple integration [9] immediately allows deriving the desired answer from Eq. (46)

$$E_2 = -\frac{9}{4}\mathcal{E}^2 .$$

## C. State l=n-1

We consider the calculation for the parameters when $l$ is maximum and equals $m$: $l = m = n-1$. The unperturbed state is the product of a homogeneous polynomial and an exponential:

$$\Psi_0 = \exp(-\frac{r}{n})(x+iy)^{(n-1)} .$$

Following the method in Sec. 4.2, we seek a function in the form

$$\Psi_1 = -(c_1 + c_2 r)\mathcal{E} z \Psi_0 .$$

Substituting into Eq.(44) yields the solution

$$\Psi_1 = -\frac{n^2}{2}(n+1+\frac{r}{n})\mathcal{E} z \Psi_0 = -\frac{n^2}{2}(n+1+\frac{r}{n})\mathcal{E} z (x+iy)^{(n-1)} \exp(-\frac{r}{n}) . \qquad (48)$$

The integrals are easy to calculate, and Eq. (46) leads to the dipole moment

$$d_z = \frac{n^4(n+1)(4n+5)}{4}\mathcal{E} , \qquad (49)$$

which coincides with Eq. (43) for $m = (n-1)$. In dimensional units, the moment is expressed as

$$d_z = \frac{n^4(n+1)(4n+5)}{4} a_B^3 \mathcal{E} .$$

An even more direct derivation of this result is possible if we note that correction Eq.(48) is equal to the combination of two states with $(n+1)$ and $(n+2)$ for $l=n$:

$$|(n+2),n\rangle = \mathcal{E} z e^{-r}(x+iy)^{(n-1)}(1-\frac{r}{(n+1)}), \qquad |(n+1),n\rangle = \mathcal{E} z e^{-r}(x+iy)^{(n-1)} .$$

The radii of the orbits are reduced to unity, and, hence, these states are not orthogonal. However, they satisfy Eq. (2) with different eigenvalues:

$$(-\Delta + 1 - \frac{2n}{r})|(n+2),n\rangle = \frac{4}{r}|(n+2),n\rangle,$$

$$(-\Delta + 1 - \frac{2n}{r})|(n+1),n\rangle = \frac{2}{r}|(n+1),n\rangle.$$

We must now compose a combination of them such that the factor $1/r$ on the right-hand side cancels, leading to the source

---

[9] The integral $\int_0^\infty e^{-r} r^k dr = k!$ accomplished by averaging Eq.(40)

$$\Psi_{n(n-1)} = (x+iy)^{(n-1)} \mathcal{E}z \exp(-r),$$

which gives

$$\frac{n^2(n+1)}{2}\Big[|(n+2),n\rangle - 2|(n+1),n\rangle\Big].$$

Substitution in the **SE** turns out to be not very complicated. (When integrating, we must return to the corresponding radii $(n+2)$, $(n+1)$ and $n$.)

## V. DIFFERENTIAL FORM of the SCHRÖDINGER EQUATION in MOMENTUM SPACE

### A. *Derivation and solution*

We apply the method of harmonic tensors to the **SE** in the momentum space. In the traditional approach, only the integral form of the **SE** is assumed [1-4, 39, 40], which is Eq. (40) or the Fock Eq. (16). The eigenfunctions then contain a factor in the form of the Gegenbauer polynomial with modified argument [39]. The second factor is a 3D tensor (or a solid spherical function) in momentum space. Direct application of the Fourier transform reduces to the Hankel transform [40], which presents no fundamental difficulty, though if specific calculations can be bulky. It is shown below that the differential equation exists, and its derivation rather simple. Its solution is found using polynomials similar in form to the 'original ones' of Eq.(7).

We proceed from Eq. (2) when the radius of the orbit (when multiplied by $n$) is reduced to unity. We multiply Eq.(2) by modulus $r$ and square operators of both sides:

$$r(-\Delta+1)r(-\Delta+1)\Psi_{nl} = 4n^2 \Psi_{nl}. \tag{50}$$

Hence,

$$\Big[r^2(\Delta-1)^2 + 2(\hat{l}_\mathbf{x}+1)(\Delta-1)\Big]\Psi_{nl} = 4n^2\Psi_{nl}, \tag{51}$$

where the "angular moment modulus" operator is $\hat{l} = (\mathbf{x}\nabla)$.

We pass on to the new function

$$\Phi_{nl} = (\Delta-1)^2 \Psi_{nl},$$

which corresponds to multiplying the spectrum by $(p^2+1)^2$ (See Fock's method in Eq. (12)). For this, we apply the operator $(\Delta-1)^2$ to Eq. (51) and swap the operators $\hat{l}_\mathbf{x}$ *and* $\Delta$. We then obtain an equation for the function $\Phi_{nl}(\mathbf{x})$:

$$\left[(\Delta-1)^2 r^2 + 2(\Delta-1)(\hat{l}_x + 3)\right]\Phi_{nl} = 4(n^2-1)\Phi_{nl}.$$

We can now pass to the spectra $a_{nl}(\mathbf{p})$ and $b_{nl}(\mathbf{p})$ by changing $\nabla_\mathbf{r} \to i\mathbf{p}$ and $\mathbf{x} \to i\nabla_\mathbf{p}$,

$$\left[-\left(\frac{(p^2+1)}{2}\right)^2 \Delta_\mathbf{p} + \frac{(p^2+1)}{2}\hat{l}_\mathbf{p}\right]b_{nl}(\mathbf{p}) = (n^2-1)b_{nl}(\mathbf{p}), \qquad (52)$$

$$b_{nl}(\mathbf{p}) = (p^2+1)^2 a_{nl}(\mathbf{p}),$$

where $\hat{l}_\mathbf{p} = (\mathbf{p}\nabla_\mathbf{p})$. Instead of integral of Eq. (10), we obtain a problem that is no more complicated than the **SE** in coordinate space, because the equation for it resembles the **SE** with the sum of two potentials. The first operator on the left-side of Eq. (52) is equal to the operator of a vector squared, whence

$$\left[-\left((p^2+1)\nabla\right)^2 + 4(p^2+1)\hat{l}_\mathbf{p}\right]b_{nl}(\mathbf{p}) = 4(n^2-1)b_{nl}(\mathbf{p}).$$

We seek a solution of the equation in the product form

$$b_{nl}(\mathbf{p}) = Y_l(\mathbf{p})\frac{1}{(p^2+1)^l} P_k\left(\frac{1}{(p^2+1)}\right), \qquad (53)$$

where $Y_l(\mathbf{p})$ is a solid spherical function (or a harmonic tensor) and the factor $P_k(u)$ is a polynomial of degree $k$.

Relation from Eq. (5) is useful when substituting. For the factor $P_k(u)$, Eq. (52) gives the second Gauss function[10]

$$_2F(\alpha,\beta,\gamma,u) = 1 + \frac{\alpha\beta}{\gamma}\frac{u}{1!} + \frac{\alpha(\alpha+1)\beta(\beta+1)}{\gamma(\gamma+1)}\frac{u}{2!} + \ldots,$$

where the parameters are

$$\alpha = -k, \quad \beta = 2l+k+2, \quad \gamma = l+\frac{3}{2}, \quad k = n-l-1, \quad u = \frac{1}{(p^2+1)}. \qquad (54)$$

We recall that the transition to the real spectrum requires multiplying the argument $\mathbf{p}$ by $n$.

Using the properties of the Gauss functions [1, 32] we can represent the solution of Eq. (53) in another form:

$$a_{nl}(\mathbf{p}) = Y_l(\mathbf{p})\frac{1}{(p^2+1)^{(n+1)}} p^{2k} {}_2F[-k,(-n+\tfrac{1}{2}),(l+\tfrac{3}{2}),-1/p^2].$$

*Example 3*

a) $n=2$, $l=0$, $\kappa=1$ (isotropic state). We double the momentum, returning to radius 2 :

---

[10] Hypergeometric function $_2F_1$

$$a_{20}(\mathbf{p}) = (const)\frac{1}{(1+4p^2)^2}(1-\frac{2}{(1+4p^2)}) \ .$$

The derivation of this formula using the Hankel transform is rather cumbersome [41].

b) $n=l+1$, $k=0$ ($l$ is maximum):

$$\Psi_{n(n-1)}(\mathbf{x}) = Y_{(n-1)}(\mathbf{x})\exp(-\frac{r}{n}) \ ,$$

where the first factor is solid spherical function, i.e., a harmonic tensor. We pass on to the argument $\mathbf{x}/n$ and apply the Fourier transform. The polynomial turns into a differential operator applied to $1/(1+p^2)^2$:

$$a_{nl}(\mathbf{p}) = \hat{Y}_l(i\nabla_\mathbf{p})\frac{1}{(1+p^2)^2} = (const)\frac{Y_l(\mathbf{p})}{(1+p^2)^{(2+l)}} \ .$$

The action of the differential operator generates the same solid spherical function (or tensor), and the next function must be differentiated $l$ times with respect to the argument $p^2$ (see Eq. (B.1) in Appendix B ).

Returning to the physical argument $n\mathbf{p}$ and using the homogeneity of the polynomial $Y_l(\mathbf{p})$, we obtain

$$a_{nl}(\mathbf{p}) = (const)\frac{Y_{(n-1)}(\mathbf{p})}{(1+n^2p^2)^{(n+1)}} \ ,$$

which coincides with the general Eq. (53).

**B.** *Group meaning of solution*

Harmonic 4-D polynomials satisfy the 4D SO(4)-invariant Laplace equation, which in the momentum Fock's space on a sphere $|\xi|^2 + \xi_0^2 = \rho_\xi^2 = 1$ has the form

$$\Delta_\xi Y_{(n-1)} = \Delta_{\perp\xi}Y_{(n-1)} + \frac{1}{\rho^3}\frac{\partial}{\partial\rho}\rho^3\frac{\partial}{\partial\rho}Y_{(n-1)} = \left[\Delta_{\perp\xi} + (n^2-1)\right]Y_{(n-1)} = 0. \tag{55}$$

We recall that degree of homogeneous polynomial is n-1 , hence , its radial dependence is $\rho_\xi^{(n-1)}$ (see foot note [5] in Section 2 ). On the sphere with $\rho_\xi = 1$, the transverse operator $\Delta_{\perp\xi}$ becomes a differential operator (with three angels), that eigenvalue is equal to $(n^2-1)$. Comparing Eqs. (52) and (53), we can see that the derived Eq. (52) after the Fock transformation of Eq. (11) with the scaling factor $(1+\mathbf{p}^2)/2$ , switches to the problem for eigenvalues of the angular part of the 4D Laplace equation on the sphere shown in the figure.

# VI. MODIFYING FOCK'S THEORY

## A. *Extension to 4-D momentum space*

Knowing the method of harmonic tensors (or solid spherical functions), we consider the inverse transition from 4D spherical functions in momentum space to the **SE** in physical space (where the radii of the orbits are reduced to unity), without repeating it literary. The main point in each transformation is to preserve the SO(4) symmetry. The first modification of the Fock theory is the transition from spherical functions to 4D solid spherical functions inside the sphere. They coincide on the surface $|\xi|^2 + \xi^2 = \rho_\xi^2 = 1$.

The starting point of the theory is therefore given by the solid spherical 4D functions

$$Y_l(\boldsymbol{\xi}) \rho_\xi^k C_k^{l+1}(\xi_0/\rho_\xi), \qquad (56)$$

where the first factor is 3D solid spherical function (or a harmonic tensor in 3D space), and the second is the Gegenbauer polynomial (with the argument $\xi_0/\rho_\xi$) multiplied by $\rho_\xi^k$. The vector properties of the Gegenbauer polynomials useful in physical problems, are discussed in Appendix E and section 6.3. Now, we only use the homogeneity property in the coordinates of the first and second factors. This polynomial is some projection of the harmonic tensor $\mathcal{M}_{[i]}^{(n-1)}(\boldsymbol{\xi}, \xi_0)$ from Eq. (32), i.e., a contraction over indices with a numerical tensor.

Instead of a stereographic projection of a sphere, we consider a closely related 4D spatial transformation consisting of three steps:

- shift of the sphere upward by one: $\xi_0' = (\xi_0 + 1)$;
- inversion ' $\dfrac{2}{\rho}$ ' of Fock's space: $(\boldsymbol{\xi}', \xi_0') = \dfrac{2}{(\mathbf{p}^2 + p_0'^2)}(\mathbf{p}, p_0')$;
- shift of the plane $p_0' = 1$ downward by one: $p_0 = p_0' - 1$.

The point $(\boldsymbol{\xi}, 0)$ on the sphere with the modulus $|\boldsymbol{\xi}| = 1$ then goes into the point $\mathbf{p} = \boldsymbol{\xi}$, which is exactly what happens under the stereographic projection. The sphere maps onto the plane. In physics, the inversion of the radius –vector relative to the sphere is more common than 'cartography' although the latter has a universal geometric description.

To come to 4D harmonic functions, it is necessary not only to replace the argument but also to divide solid spherical function by $(\mathbf{p}^2 + p_0^2)$ after inversion. We temporarily postpone the second shift of the plane along axis $p_0'$ and apply the Taylor expansion to the polynomial of degree (*n*-1) shifted along the $\xi_0$ axis by 1,

$$Y_l(\xi)C_k^{l+1}((\xi_0-1)/\tilde{\rho})\tilde{\rho}^k = Y_l(\xi)\sum_{m=1}^{n-1}\frac{(-1)^m}{m!}(\frac{\partial}{\partial \xi_0})^m\left[C_k^{l+1}(\xi_0/\rho)\rho^k\right], \quad (57)$$

where $\tilde{\rho}^2 = \xi^2 + (\xi_0-1)^2$ (the shifted squared radius). The 3D function $Y_l(\xi)$ is not affected by the shift.

We use the important property of homogeneity, which persists under the differentiation. An additional term in the sum has the homogeneity degree $n$-$1$-$m$. The factor $Y_l(\xi)$ in front of the sum, is homogeneous of degree $l$. Under inversion, the form of the functions is preserved, but inverse powers of modulus of the 4D radius vector appear. The additional factor $1/(\mathbf{p}^2 + p_0^2) = 1/\rho_\mathbf{p}^2$ turns the terms into multipole potentials:

$$\sum_{m=1}^{n-1}\frac{\mathscr{M}^{(n-1-m)}}{(\mathbf{p}^2+p_0^2)^{(n-m)}} = \sum_{m=1}^{n-1}\frac{\mathscr{M}^{(n-1-m)}}{\rho_\mathbf{p}^{2(n-m)}}. \quad (58)$$

The nominator in Eq. (58) contains solid spherical 4D functions (or 4D tensors):

$$\mathscr{M}^{(n-1-m)} = Y_l(\mathbf{p})\frac{(-1)^m 2^{(n-m)}}{m!}(\frac{\partial}{\partial p_0})^m\left[C_k^{l+1}(p_0/\rho_\mathbf{p})\rho_\mathbf{p}^k\right], \quad (59)$$

where $\rho_\mathbf{p}^2 = (\mathbf{p}^2 + p_0^2)$.

We have to recall that a shift $p_0 = p_0' - 1$ of the plane downward by the one, remains to be implemented.

### B. *Introduction of the factor $1/r$ in physical space*

It can be seen from **SE** of Eq.(2) that applying the operator $(-\Delta+1)$ to eigenfunction $\Psi_{nl}(\mathbf{x})$ is equivalent (up to a constant) to dividing it by $r$. The spectrum $a(\mathbf{p})$ is then multiplied by $(\mathbf{p}^2+1)$. We discard one factor in Eq. (12), modifying Fock's theory once again in order to work with harmonic functions. The final result, as has to be shown, is the function $\Psi_{nl}(\mathbf{x})/r$.

### C. *Transition to the 4D coordinate space*

The Fourier transformation

$$\int e^{i(\mathbf{p}\mathbf{x}+p_0\tau)}a_{nl}(\mathbf{p},p_0)d^3\mathbf{p}dp_0$$

maps the eigenfunctions $a_{nl}(\mathbf{p}, p_0)$ into the 4D coordinate space, i.e., into functions of four coordinates ($\hbar = 1$). We then apply the inverse transformation with respect to the 'extra' variable $\tau$. The reason for such a 'knight move' is that multipole tensors become harmonic ones with a factor $1/(\mathbf{x}^2 + \tau^2)$ (see Appendix C). The transformations are performed up to a 'floating' constant; we specify the normalization later.

The original function is given in Eqs. (58) and (59). The 4D Fourier transformation of the sum replaces the argument $(\mathbf{p}, p_0)$ with $(\mathbf{x}, \tau)$, changes the factorial coefficients, and adds the factor $1/(\mathbf{x}^2 + \tau^2)$. From Eq.(C.5), we then have

$$\frac{4\pi^2}{(x^2+\tau^2)} \sum_{m=1}^{n-1} \frac{i^{(n-1-m)} \mathcal{M}^{(n-1-m)}(\mathbf{x},\tau)}{2^{(n-1-m)}(n-1-m)!} . \tag{60}$$

Substituting Eq. (59) into Eq. (60), we use Newton's binomial,

$$(const)\frac{Y_l(\mathbf{x})}{(x^2+\tau^2)} \sum_{m=1}^{n-1} \frac{i^m}{(n-1-m)!} \frac{(n-1)!}{m!} (\frac{\partial}{\partial \tau})^m \left[ C_k^{l+1}(\tau/R) R^k \right] =$$
$$= (const)\frac{Y_l(\mathbf{x})}{(x^2+\tau^2)} (1+i\frac{\partial}{\partial \tau})^{(n-1)} \left[ C_k^{l+1}(\tau/R) R^k \right], \tag{61}$$

where $R = \sqrt{(\mathbf{x}^2+\tau^2)}$. Here, it is necessary to take the plain shift $p_0' = (p_0 + 1)$ into account. An extra essential factor $\exp(-i\tau)$ arises. We have thus obtained the coordinate space $(\mathbf{x}, \tau)$, where transformed eigenfunctions are proportional to $(n-1)$th derivative of solid spherical functions:

$$\tilde{\Psi}_{nl}(x,\tau) = \frac{\exp(-i\tau)Y_l(\mathbf{x})}{(\mathbf{x}^2+\tau^2)} (1+i\frac{\partial}{\partial \tau})^{(n-1)} \left[ C_k^{l+1}(\tau/R) R^k \right], \tag{62}$$

where $R^2 = (\mathbf{x}^2 + \tau^2)$.

It remains to calculate the Fourier reverse transform with respect to variable $\tau$:

$$\frac{1}{2\pi} \int \tilde{\Psi}_{nl}(\mathbf{x},\tau) e^{-i\omega\tau} d\tau . \tag{63}$$

### D. *From harmonic 4D polynomials to states of the hydrogen atom*

The Fourier transform must be understood in generalized sense, because the numerator of Eq. (62) contains a polynomial. The result is a combination of delta functions, which must be discarded. This is achieved by closing the integration contour in the lower half-plane, where the exponential decreases and the pole

$\tau = -ir$ is present:[11]

$$Y_l(\mathbf{x}) \oint_{\text{Im}\tau \leq 0} e^{-i\tau} \frac{1}{(\mathbf{x}^2 + \tau^2)} (1 + i\frac{\partial}{\partial \tau})^{(n-1)} \left[ C_k^{l+1}(\tau/R) R^k \right] d\tau . \tag{64}$$

At $\omega = 0$, the integral is therefore equal to

$$\frac{\Psi_{nl}(\mathbf{x})}{r} = (const) Y_l(\mathbf{x}) \frac{e^{-r}}{r} (1 - \frac{\partial}{\partial t})^{(n-1)} \left[ C_k^{l+1}(it/R) R^k \right]\bigg|_{t=r} . \tag{65}$$

where we recall that the polynomial $C_k^{l+1}(it)$ is either an even or an odd function. The squared 4D radius $R^2 = (r^2 + \tau^2) = (r^2 - t^2)$ is to be equated to zero only after differentiating. Each derivative with a minus sign can be pulled out of the exponential:

$$\Psi_{nl}(\mathbf{x}) = (const) Y_l(\mathbf{x}) \frac{1}{r} \frac{\partial^{(n-1)}}{\partial t^{(n-1)}} \left( e^{-t} C_k^{l+1}(it/R) R^k \right)\bigg|_{t=r} . \tag{66}$$

In deriving Eqs. (65) and (66), we deviated from Fock's theory in three points. The validity of the final result must now be verified. We take a specific state for the verification.

*Example 4*
State: $n,\ l = n - 3$.

$$\Psi_{n(n-3)} = Y_{(n-3)}(\mathbf{x}) \left( 1 - \frac{2r}{(n-2)} + \frac{2r^2}{(n-2)(2n-3)} \right),$$

$$R^2 C_2^{(n-2)}(\frac{it}{R}) = (const) \left( (it)^2 - \frac{((it)^2 + r^2)}{2(n-1)} \right),$$

$$(1 - \frac{\partial}{\partial t})^{(n-1)} = 1 - (n-1)\frac{\partial}{\partial t} + \frac{(n-1)(n-2)}{2} \left(\frac{\partial}{\partial t}\right)^2 \ldots$$

Substitution into Eq.(65) gives equality ( with a 'floating' constant ).

The general formula is verified in Appendix D, where the constant is found. For the Laguerre polynomials, we have

$$L_k^{2l+1}(r) = \frac{i^k l!}{(l+k)!} \left( (1 - \frac{\partial}{\partial t})^k R^k C_k^{l+1}(it/R) \right)\bigg|_{t=r} , \tag{67}$$

where $R^2 = (r^2 - t^2)$.

---

[11] Adding of arbitrary polynomial multiplied by an exponential doesn't change integral of Eq. (64). The polynomial can be chosen such that the delta-function does not appear in Eq. (63).

Thus, there is no need to apply Fock's transformations. It suffices to choose a harmonic polynomial (or tensor), differentiate it, and make the change $\tau = -ir$. We note here that the substitution $\tau = -it$ signifies a transition from 4D the Laplace equation to the wave equation with the wave propagation speed $Ze^2/\hbar n$ in dimensional coordinates.

## VII. DERIVATION of FOCK'S EQUATION by ELECTROSTATIC METHODS

### A. *In 3D electrostatics*

We consider the potential of charges $\sigma(\mathbf{x})$ on the surface of a unit sphere in the case where the charge density is a spherical function with index $l$ (or homogeneous harmonic polynomial of degree l):

$$\sigma(\mathbf{x}) = Y_l(\mathbf{x}) \quad (|\mathbf{x}|=1) . \tag{68}$$

This is the simplest problem of electrostatics. The potential $\varphi(\mathbf{x})$ is then outside and inside the sphere:

$$\varphi(\mathbf{x}) = c_l \frac{Y_l(\mathbf{x})}{r^{2l+1}} \quad (|r| \geq 1), \qquad \varphi(\mathbf{x}) = c_l Y_l(\mathbf{x}) \quad (|r| \leq 1), \tag{69}$$

where $Y_l(\mathbf{x})$ is a polynomial of degree $l$.

At the boundary, the potential is continuous, and the electric field has discontinuity $4\pi\sigma$ for r=1:

$$-\frac{\partial}{\partial r} c_l \frac{Y_l(\theta,\varphi)}{r^{l+1}}\bigg|_{r=1} + \frac{\partial}{\partial r} c_l Y_l(\theta,\varphi) r^l \bigg|_{r=1} = c_l(2l+1) Y_l(\theta,\varphi)\big|_{r=1} = 4\pi\sigma . \tag{70}$$

Hence,

$$c_l = \frac{4\pi}{(2l+1)} . \tag{71}$$

On the other hand, the potential is determined by Newton's integral formula

$$\int_{r'=1} \frac{\sigma(\mathbf{x}')}{|\mathbf{x}-\mathbf{x}'|} dS' = \varphi(\mathbf{x}) . \tag{72}$$

We substitute here the potential from Eq. (69) and let $\mathbf{x}$ tend to the surface. We then obtain an integral equation for 3D spherical functions,

$$Y_l(\mathbf{x}) - \frac{(2l+1)}{4\pi} \int_{r'=1} \frac{Y_l(\mathbf{x}')}{|\mathbf{x}-\mathbf{x}'|} dS' = 0 \quad (|\mathbf{x}|=1). \tag{73}$$

We use the analogy to derive Fock's Eq. (16), i.e., that for 4D spherical functions.

## B. *In 4-D space*

On the surface of a 4D sphere, we choose the charge density

$$\sigma(\mathbf{x}) = \mathcal{M}_{n-1}(\mathbf{x}) \quad (|\mathbf{x}|=1),$$

where $\mathcal{M}_{n-1}(\mathbf{x})$ is a harmonic homogeneous polynomial of degree ($n$-1) (that is any combination of 4D spherical functions with the principal number $n$-1 on the surface). Then, the potential $\varphi(\mathbf{x})$ is as follows outside and inside the sphere:

$$\varphi(\mathbf{x}) = c_{n-1} \frac{\mathcal{M}_{n-1}(\mathbf{x})}{r^{2n}} \quad (|r| \geq 1), \qquad \varphi(\mathbf{x}) = c_{n-1} \mathcal{M}_{n-1}(\mathbf{x}) \quad (|r| \leq 1). \tag{74}$$

The electric field discontinuity in this case is $4\pi^2 \sigma$ for r=1:

$$-\frac{\partial}{\partial r} c_{n-1} \frac{Y_n(\theta,\varphi,\psi)}{r^{n+1}}\bigg|_{r=1} + \frac{\partial}{\partial r} c_{n-1} Y_n(\theta,\varphi,\psi) r^{n-1}\bigg|_{r=1} = c_{n-1} 2n Y_{n-1}(\theta,\varphi,\psi)\big|_{r=1} = 4\pi^2 \sigma,$$

Hence,

$$c_{n-1} = \frac{4\pi^2}{2n}.$$

Now, instead of the Eq. (73), we obtain Fock's Eq. (16) for homogeneous harmonic polynomials of degree $(n-1)$ (see footnote 5 in Sec. II):

$$\mathcal{M}_{n-1}(\mathbf{x}) - \frac{n}{2\pi^2} \int_{r'=1} \frac{\mathcal{M}_{n-1}(\mathbf{x}')}{|\mathbf{x}-\mathbf{x}'|^2} dS' = 0 \quad (|\mathbf{x}|=1) \tag{75}$$

Thus, the equation is easier to understand in coordinate than in momentum space.

## VIII. Ladder operators in the Coulomb problem

### A. *Raising оператор $\hat{D}$*

Ladder operators are useful for representing eigenfunctions in a compact form. They are a basis for constructing coherent states [44-49]. In the Coulomb problem, they are in many respects close to the 'creation' and 'annihilation' operators of an oscillator.

The squared angular momentum operator is
$$\hat{\mathbf{L}}^2 = -r^2 \Delta + \hat{l}(\hat{l}+1),$$

where the momentum modulus operator $\hat{l}$ is equal to $(\mathbf{x}\nabla)$, become $\hat{l}(\hat{l}+1)$ when acting on the space of harmonic (or solid spherical) functions. In an invariant form, the eigenfunctions of $\hat{\mathbf{L}}^2$ are harmonic tensors of Eq. (19).

The operator $\hat{D}$ that increases the value of $l$ by one, was introduced in [26]. It can be obtained from Eq. (18):

$$\nabla_i \frac{\overset{(l)}{M_{k...m}}(\mathbf{r})}{r^{2l+1}} = -\frac{\overset{(l+1)}{M_{ik...m}}(\mathbf{r})}{r^{2l+3}} . \tag{76}$$

Straightforward differentiating of the left-hand side of Eq. (76) yields a vector operator acting on a harmonic tensor:

$$\hat{D}_i M^{(l)}_{k...m}(\mathbf{r}) = M^{(l+1)}_{ik...m}(\mathbf{r}),$$
$$\hat{\mathbf{D}} = (2\hat{l}-1)\mathbf{x} - r^2 \nabla . \tag{77}$$

In particular,

$$\hat{D}_i x_k = 3 x_i x_k - r^2 \delta_{ik} \quad , \quad \hat{D}_i \hat{D}_k \hat{D}_m 1 = \left( 3(5 x_i x_k x_m - r^2 (\delta_{ik} x_m + \delta_{ik} x_m + \delta_{ik} x_m)) \right) .$$

Applying the operator to Eq. (27) with the help of differential formula

$$\nabla_i x^{\otimes l} + x_i \ll \delta_{[..]} x^{\otimes (l-2)} \gg \; = \; \ll \delta_{[i.]} x^{\otimes (l-1)} \gg ,$$

increases the rank of the harmonic tensor.

As a result of $l$-fold application to 1, we obtain the harmonic tensor

$$\hat{D}_i \hat{D}_k ... \hat{D}_m 1 = M^{(l)}_{ik...m} = \hat{\mathbf{D}}^{\otimes l}_{[i]} 1 = \mathbf{M}^{(l)}_{[i]}, \tag{78}$$

written here in different forms.

The relation of this tensor to the angular momentum operator $\hat{\mathbf{L}}$ is as follows:

$$\hat{\mathbf{D}} = \hat{l}\mathbf{x} + i[\mathbf{x} \times \hat{\mathbf{L}}] . \tag{79}$$

As indicated in Eq. (26), when functions are inverted with respect to a sphere, $r \to 1/r$ (with an additional factor $1/r$), a harmonic tensor passes into multipole of Eq. (20), and the operator $\hat{\mathbf{D}}$ into the operator $(-\nabla)$. If we multiply the operator by $i\hbar$, then the raising operator acquires the physical meaning of a *momentum operator* in the space inverted with respect to the sphere. The lowering operator is obvious: it is gradient. Under inversion, it becomes $(-\hat{\mathbf{D}})$.

We give some useful properties of operators in vector form:

$$\hat{D}_i \hat{D}_i = r^4 \Delta , \tag{80}$$

which for tensors yields a vanishing trace over any two indices. The scalar products of vectors $\hat{\mathbf{D}}$ and $\mathbf{x}$ are

$$(\mathbf{x}\hat{\mathbf{D}}) = (\hat{l}+1)r^2 , \quad (\hat{\mathbf{D}}\mathbf{x}) = r^2 \hat{l} . \tag{81}$$

Hence, the contraction of the tensor with the vector $\mathbf{x}$ can be expected as

$$x_i M^{(l)}_{ik...m} = l\, r^2 M^{(l-1)}_{k...m}, \tag{82}$$

where $l$ is a number.

The commutator in the scalar product on the sphere is equal to unity:

$$(\mathbf{x}\hat{\mathbf{D}}) - (\hat{\mathbf{D}}\mathbf{x}) = r^2.$$

To calculate the divergence of a tensor, a useful formula is
$$\nabla \mathbf{D} = (\hat{l}+1)(2\hat{l}+3),$$
whence
$$\nabla_i M^{(l)}_{ik...m} = l(2l+1)M^{(l-1)}_{k...m}. \tag{83}$$
($l$ on the right-hand side is a number).

The raising operator is useful for constructing the factor $Y_l(\mathbf{x})$ in front of the radial function in the solution structure of SE of Eq. (7).

## B. *Four-dimensional operator* $\hat{\mathcal{D}}$

The raising operator in 4-D space
$$\hat{\mathcal{D}}_i M^{(n)}_{k...m}(\mathbf{r},\tau) = M^{(n+1)}_{ik...m}(\mathbf{r},\tau) = M^{(n+1)}_{ik...m}(\mathbf{y}), \tag{84}$$
has largely similar properties. There is a raising operator
$$\hat{\mathcal{D}} = 2\hat{n}\mathbf{y} - |\mathbf{y}|^2 \nabla_\mathbf{y},$$
where $y_i$ is 4D vector, $i=1,...4$,
$$\mathbf{y} = (\mathbf{x},\tau), \qquad \rho_\mathbf{y}^2 = (\mathbf{x}^2 + \tau^2),$$
and the $\hat{n}$ operator multiplies a homogeneous polynomial by its degree[12],
$$\hat{n} = (\mathbf{x}\nabla_\mathbf{x} + \tau\frac{\partial}{\partial \tau}) = (\mathbf{y}\nabla_\mathbf{y}), \tag{85}$$
In particular,
$$\hat{\mathcal{D}}_i 1 = 2y_i, \quad \hat{\mathcal{D}}_i y_k = 2(4 y_i y_k - \rho_\mathbf{y}^2 \delta_{ik}),$$
$$\hat{\mathcal{D}}_i \hat{\mathcal{D}}_k \hat{\mathcal{D}}_m 1 = 4!!\left((6 x_i x_k x_m - \rho_\mathbf{y}^2(\delta_{ik}x_m + \delta_{ik}x_m + \delta_{ik}x_m))\right).$$

The scalar product of $\hat{\mathcal{D}}$ and $\mathbf{y}$ is as simple as in Eq. (81)
$$(\mathbf{y}\hat{\mathcal{D}}) = \hat{n}\rho_\mathbf{y}^2, \qquad (\hat{\mathcal{D}}\mathbf{y}) = (\hat{n}-2)\rho_\mathbf{y}^2.$$

The scalar product of ladder operators $\hat{\mathcal{D}}$ and $\nabla_\mathbf{y}$ is
$$(\nabla\hat{\mathcal{D}}) = 2(\hat{n}+2)^2.$$

The transformed solution of the **SE** can be regarded as an invariant tensor in the 4-D coordinate space $\mathbf{y}$. It is then convenient to use the $\hat{\mathcal{D}}$ raising operator.

As in Eq. (79), operator $\hat{\mathcal{D}}$ is now associated with the angular momentum operator and the Laplace-Runge-Lenz operator $\hat{\mathbf{A}}$, which in the 4D coordinate space takes the simple form

---

[12] Separating the $\tau$ variable is necessary in order to pass to the physical space.

$$\hat{\mathbf{A}} = i(\tau \frac{\partial}{\partial \mathbf{x}} - \mathbf{x}\frac{\partial}{\partial \tau}) \ .$$

Separately, for 3D $\mathbf{x}$ -component and the fourth coordinate $\tau$ of the raising operator, we have

$$\hat{\mathcal{D}}_{\mathbf{x}} = (\hat{n}+1)\mathbf{x} + i[\mathbf{x}\times\hat{\mathbf{L}}] + i\tau\mathbf{A} \ ,$$

$$\hat{\mathcal{D}}_{\tau} = (\hat{n}+1)\tau + i(\mathbf{xA}) \ .$$

## IX. The Schwinger resolvent

Follow J. Schwinger, we consider perturbed Fock's Eq. (16) in the 4D coordinate space on a sphere, using the unit 4D vectors $\mathbf{x}, \mathbf{y}$. When perturbation is a delta function on the sphere, the solution is called resolvent $G(\mathbf{x},\mathbf{y})$. Fock's equation thus turns into an equation for the resolvent,

$$G(\mathbf{x},\mathbf{y}) - \frac{\lambda}{2\pi^2}\int_{|\mathbf{y}'|=1}\frac{G(\mathbf{xy}')}{|\mathbf{y}-\mathbf{y}'|^2}dS_{\mathbf{y}'} = \delta(\mathbf{x}-\mathbf{y}) \ , \qquad (86)$$

where instead of $n$ we have a continuous parameter $\lambda$.

    J. Schwinger applied the entire series of the integral perturbation theory to the solution of Eq. (86) in order to show that some series in quantum electrodynamics (**QED**) are summable [50]. This assumption is confirmed in some cases [56]. For the Coulomb problem, the answer is expressed in terms of elementary functions.

    The solution of this problem can be obtained by a simple calculation [3]. of the Gegenbauer polynomials with the properties listed in Appendix E [51-53] We seek a solution of Eq. (86) in the form of a sum of polynomials. We use the obvious expansions

$$\frac{1}{|\mathbf{x}-\mathbf{y}|^2} = \sum_{k=0}^{\infty} C_k^1(\mathbf{xy}) \ , \qquad (87)$$

$$\frac{1}{2\pi^2}\sum_{k=0}^{\infty}(k+1)C_k^1(\mathbf{xy}) = \delta(\mathbf{x}-\mathbf{y}) \ , \qquad (88)$$

which immediately gives a simple series

$$G(\mathbf{x},\mathbf{y}) = \frac{1}{2\pi^2}\sum_{k=0}^{\infty} C_k^1(\mathbf{xy})\frac{(k+1)^2}{(k+1-\lambda)} \ . \qquad (89)$$

To improve convergence, we subtract series of Eq. (87) multiplied by the coefficient $\lambda/2\pi^2$ from it, and then series of Eq. (88). We thus obtain [13]

$$G(\mathbf{x},\mathbf{y}) = \delta(\mathbf{x}-\mathbf{y}) + \frac{\lambda}{2\pi^2|\mathbf{x}-\mathbf{y}|^2} + \frac{\lambda^2}{2\pi^2}\sum_{k=0}^{\infty}C_k^1(\mathbf{xy})\frac{1}{(k+1-\lambda)} \ . \qquad (90)$$

---

[13] The sign of $\lambda$ is changed in [3].

This is already a useful result for calculations regarding the vector properties of the Gegenbauer polynomials[14].

The resolvent has poles at $\lambda = 1, 2...$, which corresponds to polynomials with the numbers $k = 0, 1, 2....$. The values of the main quantum number $n$ in the **SE** are 1, 2, ..., respectively.

## X.  DISCUSSION

We illustrate the result of this study with the ground state of **SE** ($n = 1$, $\Psi_1 = \exp(-r)$). Passing to the momentum description gives the function $1/(p^2+1)^2$. Fock's transformation gives the state with the number $n-1=0$ *on the surface of a sphere*, i.e., function 1. In the coordinate 4D space, we must take the harmonic polynomial with the number 0 *in the ball*, i.e., the polynomial 1, and then multiply the unit by $\exp(-r)$. This brings us back to the physical space without integrals.

The general formula for the transformation from the 4D space is just as simple. In any solution of the Laplace equation given by a homogeneous polynomial of degree $n-1$, including in the form of invariant tensor of Eq. (32), we replace the coordinate $\tau = -it$. Then, we multiply the polynomial by $\exp(-t)$ and calculate the $(n-1)$th derivative. The last step is to equate $t = r$ and obtain the n state.

To consider states with a fixed value of $l$, we must take a harmonic 3D polynomial of degree $l$ (or 3D tensor of Eq. (27)) in the 3D coordinate space and then compose a polynomial of the form from Eq. (66), which includes the Gegenbauer polynomial, and carry out a simple recalculation as in Eq. (65).

We note that, although it is no longer necessary to apply the Fock transformation, some results obtained using 4D spherical functions, carry over to the time-coordinate 4D space, including the Schwinger resolvent and the Fock integral equation. It is only necessary to replace the 'momentum' arguments $(\xi, \xi_0)$ with the time- coordinate vector $(\mathbf{x}, \tau)$.

As regards the method, we can see that solutions of the 3D and 4D Laplace equations play an important role, and tensor invariant methods of electrostatics are applicable. Importantly, the replacement $\tau = -it$ means the transition to the wave equation with the wave propagation speed $Ze^2/\hbar n$. Then, the time parameter is set equal to the radius and disappears. The space in which the SO(4) symmetry is realized turns out to be 'closer' to the physical one than does the wrapped momentum space.

---

[14] J. Schwinger sums the series by introducing a parameter $\zeta$ in the integral: $\int_0^1 \zeta^{k+1-\lambda} d\zeta$

**Acknowledgements**

The author would like to express profound gratitude to Prof. D. Bohm for stimulating discussion of methods of quantum theory in 1991 y. The author is deeply grateful to Prof. B. Bolotovskii for his attention to this study and his useful remarks.

# XI. Appendices

## A. Structure of harmonic tensor

We apply the operator $\Delta$ to the sum of tensors in Eq. (24). Using the properties of of homogeneous polynomials in Eq. (5), we obtain two terms for the $k$ th term of the sum:

$$\Delta\left[(-1)^k \frac{(2l-2k-1)!!}{k!2^k} r^{2k} \Delta^k \mathbf{r}^{\otimes l}\right] = \left[(-1)^k \frac{(2l-2k-1)!!}{k!2^k} r^{2k} \Delta^{(k+1)} \mathbf{r}^{\otimes l}\right] + \\ +2k(2l-2k+1)\left[(-1)^k \frac{(2l-2k-1)!!}{k!2^k} r^{2k-2} \Delta^k \mathbf{r}^{\otimes l}\right] \qquad (A.1)$$

We next take the second term of the sum, increasing number $k$ by one:

$$2(k+1)\left[(-1)^{k+1} \frac{[2l-2(k+1)+1]!!}{(k+1)!2^{(k+1)}} r^{2k} \Delta^{(k+1)} \mathbf{r}^{\otimes l} = \right] \\ = -\left[(-1)^k \frac{[2l-2k-1]!!}{(k)!2^k} r^{2k} \Delta^{(k+1)} \mathbf{r}^{\otimes l}\right] \qquad (A.2)$$

This term 'absorbs' the first one from the term of the previous sum. As a result, the total sum is equal to zero and tensor of Eq. (24) is a harmonic function.

## B. Harmonic tensor as an operator

The Fourier transform maps a polynomial into a differential one. An invariant harmonic tensor is mapped into a tensor operator in the momentum space:

$$\mathbf{M}_{[\mathbf{i}]}^{(l)}(\mathbf{x}) \to i^l \hat{\mathbf{M}}_{[\mathbf{i}]}^{(l)}(\nabla_{\mathbf{p}}).$$

We consider how it acts on a scalar function of the argument $\mathbf{p}^2 = p^2$. The result is quite obvious and has been 'rediscovered' several times [54]. Presumably, it was already known to Hobson[19]. We propose the following derivation.

From uniqueness of invariant homogeneous tensor that is homogeneous in coordinates, it follows that the result is proportional to the harmonic tensor:

$$\hat{\mathbf{M}}_{[\mathbf{i}]}^{(l)}(\nabla_{\mathbf{p}}) f(p^2) = \mathbf{M}_{[\mathbf{i}]}^{(l)}(\mathbf{p}) \varphi(p).$$

The unknown function $\varphi(p)$ can be found by taking a power of complex variable instead of a tensor: $(x+iy)^l = u^l$. The operator arises raised to the power $l$:

$$\left(\frac{\partial}{\partial x} + i\frac{\partial}{\partial y}\right)^l = 2^l \frac{\partial^l}{\partial \bar{u}^l}.$$

This operator is applied to the function $\varphi(u\bar{u})$, with no differentiation with respect to $u$. The result is obvious:

$$2^l \frac{\partial^l}{\partial \bar{u}^l} f(u\bar{u}) = 2^l u^l f^{(l)}(u\bar{u}) = 2^l (x+iy)^l f^{(l)}(u\bar{u}),$$

where $f^{(l)}(u\bar{u})$ is the $l$th derivative. Thus, a simple and convenient formula is valid:

$$\hat{\mathbf{M}}_{[\mathbf{i}]}^{(l)}(\nabla_{\mathbf{p}}) f(p^2) = \mathbf{M}_{[\mathbf{i}]}^{(l)}(\mathbf{p}) 2^l f^{(l)}(p^2). \tag{B.1}$$

For the function $f(r^2) = 1/|\mathbf{x}|$, this formula yields the relation

$$\hat{\mathbf{M}}_{[\mathbf{i}]}^{(l)}(\nabla_{\mathbf{x}}) \frac{1}{\sqrt{|\mathbf{x}^2|}} = (-1)^l (2l+1)!! \frac{\mathbf{M}_{[\mathbf{i}]}^{(l)}(\mathbf{x})}{r^{(2l+1)}}, \tag{B.2}$$

which can be usefully compared with the definition of tensor of Eq. (18).

## C.  The 4D Fourier transform of multipole potentials

The Fourier transform of multipole potentials

$$\int e^{i(\mathbf{px}+p_0\tau)} \frac{\mathcal{M}_{[\mathbf{i}]}^{(n)}(\mathbf{p}, p_0)}{(\mathbf{p}^2 + p_0^2)^{(n+1)}} d^3\mathbf{p}\, dp_0, \tag{C.1}$$

preserves their form up to a scalar factor $1/(\mathbf{x}^2 + \tau^2)$, which, as shown below, follows from the Poisson equation.

We use the definition of multipole potentials in terms of 4-D gradients of Eq.(28):

$$\frac{\mathcal{M}_{i...k}^{(n)}(\mathbf{x}, \tau)}{(\mathbf{x}^2 + \tau^2)^{(n+1)}} = (-1)^n \nabla_i ... \nabla_k \frac{1}{(\mathbf{x}^2 + \tau^2)} = \frac{(-1)^n}{2^n n!} \hat{\mathcal{M}}_{i...k}^{(n)}(\nabla) \frac{1}{(\mathbf{x}^2 + \tau^2)}, \tag{C.2}$$

where the argument on the right-hand side involves the 4-D gradient. The indices $i,...,k$ are four-dimensional. This replacement is valid because all terms in Eq. (28) except the first one, contain the 4D Laplace operator (excluding the point 0).

We proceed from the equation for a point charge in 4D space:

$$(\Delta_{\mathbf{p}} + \frac{\partial^2}{\partial p_0^2}) \frac{1}{(\mathbf{p}^2 + p_0^2)} = -4\pi^2 \delta(\mathbf{p})\delta(p_0) \ . \tag{C.3}$$

We apply the operator $\hat{\mathcal{M}}^{(n)}_{i...k}(\nabla)$ to (C.3):

$$(\Delta_{\mathbf{p}} + \frac{\partial^2}{\partial p_0^2}) \frac{\mathcal{M}^{(n)}_{i...k}}{(\mathbf{p}^2 + p_0^2)^{(n+1)}} = -\frac{4\pi^2 (-1)^n}{2^n n!} \hat{\mathcal{M}}^{(n)}_{i...k}(\nabla_{\mathbf{p}})\delta(\mathbf{p})\delta(p_0) \ . \tag{C.4}$$

The trace over any two indices on the right- and left-hand sides is equal to zero.

The Fourier transform of Eq. (C.1) can now be applied to Eq. (C.4) by substituting $\nabla_{\mathbf{p}} \to -i(\mathbf{x}, \tau)$. The delta function pass to unity and we have

$$\int \exp(i(\mathbf{p}\mathbf{x} + p_0 \tau)) \frac{\mathcal{M}^{(n)}_{i...k}(\mathbf{p}, p_0)}{(\mathbf{p}^2 + p_0^2)^{(n+1)}} d^3\mathbf{p} dp_0 = \frac{4\pi^2 i^n}{2^n n!} \frac{\mathcal{M}^{(n)}_{i...k}(\mathbf{x}, \tau)}{(\mathbf{x}^2 + \tau^2)} \ . \tag{C.5}$$

The last relation holds for tensor projection, i.e., contraction over indices with a numerical tensor and, hence, *for any solid spherical function.*

**D. Polynomial correspondence**

To verify Eq. (65), we cancel $\frac{\exp(-r)}{r}$ and the spherical function $Y_l(\mathbf{x})$ (or the harmonic tensor) to have

$$F(-k, (2l+2), 2r) = (const)(1 - \frac{\partial}{\partial t})^{(l+k)} \left[ C_k^{l+1}(it/R) R^k \right]_{t=r} \ , \tag{D.1}$$

where $R^2 = (r^2 - t^2)$.

The general form of the Gegenbauer polynomial $C_k^{l+1}(t)$ is not needed. It suffices to know two of its characteristics:

$$C_k^{l+1}(1) = \frac{(2l+1+k)!}{k!(2l+1)!} \ , \tag{D.2}$$

$$C_k^{l+1}(t) = \frac{2^k (k+l)!}{k! l!} t^k + \dots \ . \tag{D.3}$$

We note that the derivative with respect to $t$ translates the 4-D harmonic function into a function whose value $k$ has decreased by one while maintaining $l$:

$$Y_l(\mathbf{x}) \frac{\partial}{\partial t} \left[ C_k^{l+1}(t/R) R^k \right] = c_{lk} Y_l(\mathbf{x}) C_{k-1}^{l+1}(t/R) R^{k-1} \ . \tag{D.4}$$

Equating $r=0$, pulling out $t^k$ and differentiating with the use of Eq. (D.2), we find the constant $c_{lk}$:

$$c_{lk} = (2l+k+1).$$

Hence, the *m*th derivative is

$$\frac{\partial^m}{\partial t^m}\left[C_k^{l+1}(t/R)R^k\right] = \frac{(2l+k+1)!}{(2l+k+1-m)!}C_{k-1}^{l+1}(t/R)R^{k-1}. \tag{D.5}$$

The function $\left[C_k^{l+1}(it/R)R^k\right]_{t=r}$ doesn't contain a power of $R$ only in the leading term; for $R=0$, therefore, the leading coefficients from Eq. (D.3) arise in Eq. (D.1) with different $k$:

$$\left[C_k^{l+1}(it/R)R^k\right]_{t=r} = \frac{2^k(k+l)!}{k!\,l!}(ir)^k. \tag{D.6}$$

On the right-hand side of Eq. (D.1), we have

$$\sum_m \frac{(l+k)!(-i)^m}{m!(l+k-m)!} \frac{(2l+k+1)!}{(2l+k+1-m)!} \frac{2^{k-m}(k-m+l)!}{(k-m)!\,l!}(ir)^{k-m}. \tag{D.7}$$

We change $k$ to $(k-m)$ and, pulling out a factor, obtain

$$\frac{(-i)^k(l+k)!(2l+k+1)!}{l!}\sum_m \frac{(-1)^m(2r)^m}{m!(k-m)!}\frac{1}{(2l+m+1)!}. \tag{D.8}$$

The sum is to be compared with the Gauss function of Eq. (6):

$$F(-k,2l+2,2r) = k!(2l+1)!\sum_m \frac{(-1)^m(2r)^m}{m!(k-m)!}\frac{1}{(2l+m+1)!}(ir)^{k-m}. \tag{D.9}$$

We arrive at a correspondence and derive the constant.

For the Laguerre polynomials, this correspondence becomes

$$L_k^{2l+1}(r) = \frac{i^k l!}{(l+k)!}(1-\frac{\partial}{\partial t})^k\left(R^k C_k^{l+1}(it/R)\right)\bigg|_{t=r}, \tag{D.10}$$

where $R^2 = (r^2 - t^2)$.

We note that Eq. (D.10) is not known in the theory of special functions [32]. For the perturbation theory, it is useful to recalculate the normalized functions. The norm in the physical space is

$$\int |Y_l|^2 dS_2 \int_0^\infty \left[L_k^{2l+1}(r)\right]^2 r^{2l+2} e^{-2r} dr = N_L^2, \tag{D.11}$$

where $S_2$ is a sphere of unit radius; the harmonic functions are normalized on the 3D sphere:

$$\int |Y_l|^2 dS_2 \frac{2}{\pi}\int_{-1}^1 \left[C_k^{l+1}(r)\right]^2 (1-t^2)^{l+1/2} dt = N_C^2. \tag{D.12}$$

The relation between the norms is given by

$$N_L = \frac{(n+l)!(n-1)!}{2} N_C. \tag{D.13}$$

We recall that the radius of the orbit is assumed to be unity. When passing to the radius equal to $n$, norm from Eq. (D.11) must be multiplied by $n^{l+3/2}$.

### E. Vector properties of the Gegenbauer polynomials

4-D the Laplace equation is simpler than 3-D one. This can be seen from the structure of harmonic tensor of Eq. (31). To construct harmonic polynomials, a simple but important property is used. Any scalar function of the form

$$\frac{1}{r} f(\tau \pm ir) ,\tag{E.1}$$

where $r = \sqrt{(x_1^2 + x_2^2 + x_3^2)}$ is 3-D radius, $\tau$ is the forth coordinate, is a harmonic function. The verification is obvious

$$\left(\frac{\partial^2}{\partial \tau^2} + \frac{1}{r}\frac{\partial^2}{\partial r^2} r \right)\frac{1}{r} f(r \pm i\tau) = 0 .$$

This gives rise to a harmonic homogeneous polynomial of degree $k$ of four variables:

$$\frac{1}{2ir}\left((\tau + ir)^{k+1} - (\tau - ir)^{k+1}\right) ,\tag{E.2}$$

which Gegenbauer writes in the form

$$\frac{1}{2ir}\left((\tau + ir)^{k+1} - (\tau - ir)^{k+1}\right) = (r^2 + \tau^2)^k C_k^1\left(\frac{\tau}{(r^2 + \tau^2)}\right),\tag{E.3}$$

where $C_k^1(z)$ is nonhomogeneous (but definite- parity) polynomial named after him. On the surface of a sphere, where $r^2 + \tau^2 = 1$, Eq. (E.3) leads to the simple form of the harmonic polynomial:

$$C_k^1(\cos\varphi) = \frac{1}{2i\sin\varphi}\left(e^{i\varphi(k+1)} - e^{-i\varphi(k+1)}\right) = \frac{\sin(k+1)\varphi}{\sin\varphi} .\tag{E.4}$$

where

$$\cos\varphi = \tau, \quad \sin\varphi = r, \quad r^2 + \tau^2 = 1.$$

Under rotations of the coordinate system, when the vector $(0,0,0,1)$ goes into a 4vector $\mathbf{x}'$ and the vector $(x_1, x_2, x_3, \tau)$ passes into $\mathbf{y}$, a polynomial arises from the scalar product of unit vectors [55]:

$$C_k^1(\mathbf{xy}) = \frac{\sin(k+1)\varphi}{\sin\varphi} , \quad \cos\varphi = (\mathbf{xy}), \quad (\mathbf{x}^2 = 1, \; \mathbf{y}^2 = 1) .\tag{E.5}$$

where the prime at $\mathbf{x}$ (') is omitted. This is the value of the harmonic function in both variables $\mathbf{x}$ and $\mathbf{y}$ in space transferred to the surface of the unit sphere.

The polynomials $C_k^1(\mathbf{x},\mathbf{y})$ play an important role in solving 4-D the Laplace equation (and 4D electrostatics): they are similar to the Legendre polynomials in 3D space. We list the properties that are invariant in the SO(4) symmetry.

Decomposition of kernel on a sphere:

$$\frac{1}{|\mathbf{x}-\mathbf{y}|^2} = \sum_{k=0}^{\infty} C_k^1(\mathbf{xy}) \ . \tag{E.6}$$

- Completeness of the system, i.e., the delta function decomposition:

$$\frac{1}{2\pi^2}\sum_{k=0}^{\infty}(k+1)C_k^1(\mathbf{xy}) = \delta(\mathbf{x}-\mathbf{y}) \ . \tag{E.7}$$

- The eigenvalue problem (see Fock's Eq. (16), now with $C_{n-1}^1$) :

$$\int_{|\mathbf{y}'|=1}\frac{C_k^1(\mathbf{xy}')}{|\mathbf{y}-\mathbf{y}'|^2}dS_{\mathbf{y}'} = \frac{2\pi^2}{(k+1)}C_k^1(\mathbf{xy}) \ . \tag{E.8}$$

- An important evaluation: $C_k^1(1) = (k+1)$ . $\tag{E.9}$

- Orthogonality and normalization:

$$\int_{|\mathbf{y}'|=1} C_k^1(\mathbf{xy}')C_l^1(\mathbf{y}'\mathbf{y})dS_{\mathbf{y}'} = \frac{2\pi^2}{(k+1)}C_k^1(\mathbf{xy})\delta_{kl} \ . \tag{E.10}$$

- Almost obvious addition theorem on the surface of a sphere [56]:

$$\frac{2\pi^2}{(k+1)}\sum_{l,m}Y_{klm}(\mathbf{x})Y_{klm}(\mathbf{y}) = C_k^1(\mathbf{xy}) \ . \tag{E.11}$$

- Leading term of the expansion of the polynomial in powers:

$$C_k^1(\mathbf{xy}) = \left(2(\mathbf{xy})\right)^k + ... \tag{E.12}$$